\documentstyle[12pt,epsf]{article}
\def\slashchar#1{\setbox0=\hbox{$#1$}           
   \dimen0=\wd0                                 
   \setbox1=\hbox{/} \dimen1=\wd1               
   \ifdim\dimen0>\dimen1                        
      \rlap{\hbox to \dimen0{\hfil/\hfil}}      
      #1                                        
   \else                                        
      \rlap{\hbox to \dimen1{\hfil$#1$\hfil}}   
      /                                         
   \fi}                                         %
\renewcommand{\theequation}{\arabic{section}.\arabic{equation}}
\def\Journal#1#2#3#4{{#1} {\bf #2}, #3 (#4)}

\def\NP{\em Nucl.~Phys.}
\def\PLB{{\em Phys.~Lett.}~B}

\def\PRD{{\em Phys.~Rev.}~D}

\def\mEt{\mbox{${\hbox{$E$\kern-0.6em\lower-.1ex\hbox{/}}}_T$}\, } 
\def\cbeta{c_\beta}
\def\sbeta{s_\beta}
\def\cW{c_W}
\def\sW{s_W}

\def\mZ{m_Z}

\def\hatu{\hat u}
\def\hatt{\hat t}
\def\hats{\hat s}
\def\stilde{\widetilde}
\def\NI{\stilde N_1}

\newcommand{\newc}{\newcommand}
\newc{\lcal}{\int {\cal L}dt}
\newc{\gsim}{\lower.7ex\hbox{$\;\stackrel{\textstyle>}{\sim}\;$}}
\newc{\lsim}{\lower.7ex\hbox{$\;\stackrel{\textstyle<}{\sim}\;$}}
\def\beq{\begin{eqnarray}}
\def\eeq{\end{eqnarray}}
\def\bea{\begin{eqnarray}}
\def\eea{\end{eqnarray}}
\catcode`@=11
\long\def\@caption#1[#2]#3{\par\addcontentsline{\csname
  ext@#1\endcsname}{#1}{\protect\numberline{\csname
  the#1\endcsname}{\ignorespaces #2}}\begingroup
    \small
    \@parboxrestore
    \@makecaption{\csname fnum@#1\endcsname}{\ignorespaces #3}\par
  \endgroup}
\catcode`@=12

\begin{document}
\begin{titlepage}
\begin{flushright}
hep-ph/0005116 \\
FERMILAB-Pub-00/097-T
\end{flushright}
\vspace{0.3in}

\baselineskip=20pt
\begin{center}
{\large\bf
Collider signals from slow decays in
supersymmetric models with an intermediate-scale
solution to the $\mu$ problem}

\end{center}

\vspace{.15in}
\begin{center}

{\sc Stephen P.~Martin}

\vspace{.1in}
{\it Department of Physics, Northern Illinois University, DeKalb IL 60115
{$\rm and$}\\
}{\it Fermi National Accelerator Laboratory,
P.O. Box 500, Batavia IL 60510 \\} 
\end{center}

\vspace{0.8in}

\begin{abstract}\noindent The problem of the origin of the $\mu$ parameter
in the Minimal Supersymmetric Standard Model can be solved by introducing
singlet supermultiplets with non-renormalizable couplings to the ordinary
Higgs supermultiplets. The Peccei-Quinn symmetry is broken at a scale
which is the geometric mean between the weak scale and the Planck scale,
yielding a $\mu$ term of the right order of magnitude and an invisible
axion. These models also predict one or more singlet fermions which have
electroweak-scale masses and suppressed couplings to MSSM states. I
consider the case that such a singlet fermion, containing the axino as an
admixture, is the lightest supersymmetric particle.  I work out the
relevant couplings in several of the simplest models of this type, and
compute the partial decay widths of the next-to-lightest supersymmetric
particle involving leptons or jets. Although these decays will have an
average proper decay length which is most likely much larger than a
typical collider detector, they can occasionally occur within the
detector, providing a striking signal.  With a large sample of
supersymmetric events, there will be an opportunity to observe these
decays, and so gain direct information about physics at very high energy
scales.

\end{abstract}
\end{titlepage}
\baselineskip=17.4pt
\setcounter{footnote}{1}
\setcounter{page}{2}
\setcounter{figure}{0}
\setcounter{table}{0}

\tableofcontents
\section{Introduction}
\label{sec:intro}
\setcounter{equation}{0}
\setcounter{footnote}{1}

Supersymmetry (for reviews, see \cite{HaberKane,primer}) has the ability
to
stabilize the hierarchy between the electroweak and Planck scales.
However, the minimal supersymmetric standard model (MSSM) still requires
an explanation for the magnitude of the supersymmetric Higgs mass
parameter $\mu$. Assuming that there are no fine-tuned cancellations in
the MSSM Higgs potential, $\mu$ should be of roughly the same magnitude as
the soft supersymmetry-breaking masses. This suggests that $\mu$ arises as
a vacuum expectation value (VEV) which is fixed by a potential with
dimensionful parameters that are in turn determined by supersymmetry
breaking.

Supersymmetry also requires some additional structure in order to solve
the strong CP-problem. The $\mu$ parameter breaks the Peccei-Quinn
(PQ) symmetry \cite{PQ} that is otherwise naturally present in the MSSM at
the
renormalizable level, so it is an attractive proposition that the dynamics
which leads to the $\mu$ term simultaneously provide for an invisible
axion.
Astrophysical constraints on the axion leave open a window 
\cite{axionwindow} from roughly
\beq 
10^{9}\>{\rm  GeV} \lsim f \lsim 10^{12}\>{\rm  GeV}
\label{axionwin}
\eeq
for the VEV of the PQ-breaking field. 

In this paper I will consider the
phenomenology of a class of models with an invisible axion
\cite{KSVZ,DFSZ} of the DFSZ type \cite{DFSZ,kimnilles,Murayama:1992dj}.
In these models, the $\mu$ term arises from non-renormalizable terms
in the superpotential, for example:
\beq 
W = {\lambda_\mu \over M_P} X_1 X_2 H_u H_d .
\label{Waxion}
\eeq
(Here $M_P = 2.4
\times 10^{18}$ GeV is the reduced Planck mass, and $\lambda_\mu$ is a
dimensionless coupling which I assume is not much smaller than unity.)
The sum of the PQ charges of $X_1$ and $X_2$ must be equal and opposite
to that of the MSSM Higgs superfields.
When the scalar components of the neutral chiral superfields $X_1$ and
$X_2$ acquire VEVs of order $f$, the approximate PQ symmetry is
spontaneously broken,
giving rise to a $\mu$ term of the right order of magnitude and an
invisible axion. It is natural to assume that this occurs with $X_1$ and
$X_2$ along a nearly flat direction in the potential.
Then the low-energy degrees of freedom will typically
include a pair of neutral chiral supermultiplets which are mixtures
of the original $X_1$ and $X_2$. One of these 
contains as its imaginary scalar component the invisible axion of the
model, and its fermion superpartner, the ``axino". 
In some models, $X_1$ and $X_2$ are the same field, so that the axino is
the only new light singlet fermion. 
If $X_1$ and $X_2$ are distinct, then there will be another light
Majorana fermion ``singlino". 
The axino and singlino  both have odd $R$-parity and can mix.
They obtain masses which are not much larger than the
weak scale (but
might be as small as of order a keV depending on the details of
the model \cite{axinomass}). 
The upper limit can be understood from the facts that the
(mass)$^2$ splittings between members of the same supermultiplet are
bounded above by roughly $m^2_{3/2}$, the squared gravitino mass,
and the axion is nearly massless. 

The axino and the singlino are very weakly coupled to MSSM states, and
cannot be directly produced in collider experiments at any significant
rate. However, if either (or both) of these particles is lighter than all
of the MSSM superpartners, then it will be the lightest supersymmetric
particle (LSP) and can appear in decays from ordinary supersymmetric
events.\footnote{If $R$-parity is conserved and the singlino or the axino
is the LSP, it will be
absolutely stable
and could dominate the energy density of the universe too soon. This
potential problem can be solved (as in many similar cases) by invoking a
low reheat
temperature, at the cost of requiring in addition a low-scale baryogenesis
mechanism. Furthermore, the NLSP decays will safely occur
long before nucleosynthesis in the standard cosmology.} 
In this paper I will argue there will be an opportunity in the
era after supersymmetry is discovered to search for and measure the very
long
lifetime of the next-to-lightest supersymmetric particle (NLSP) into final
states that include the singlino or axino, despite its very weak coupling.

In order to discuss the phenomenology in a general way, I will
denote the relevant axino or singlino LSP by $\stilde S$, and refer to
it generically as a singlino. It
is part of a
superfield $S$.
In the low-energy theory it participates in a superpotential
of the form
\beq
W = \mu \left (1 + {\epsilon \over v} S\right ) H_u H_d 
+ {1\over 2} m_{\stilde S} S^2 
\label{Wlight}
\eeq
in which 
\beq
\mu = {\lambda_\mu \over M_P}\langle X_1 \rangle \langle X_2 \rangle ,
\eeq
and I have introduced a dimensionless coupling parameter 
\beq
\epsilon \sim
v/f
\eeq with $v = 175$ GeV, the electroweak scale. 
(There are also soft mass terms for the scalar components of $S$, which
will not concern us.) For example, if $X_1$ and $X_2$ are the same
field, then one can read off from eq.~(\ref{Waxion}) that
$\epsilon = 2 v/f $.
For numerical purposes, this paper will use as a benchmark the value
$\epsilon \approx
10^{-8}$ corresponding to 
$f \approx$ (few)$\times 10^{10}$ GeV. 
There is also a dimensionless, holomorphic soft term in the lagrangian
\beq
-{\cal L}_{{\rm SUSY}\>{\rm breaking}} = {h_b \over M_P} X_1 X_2 H_u H_d
+ {\rm c.c.}
\eeq
where $h_b$ is of order $m_W$. This gives rise to (among other terms) the
necessary holomorphic soft
(mass)$^2$ term
for the Higgs bosons in the MSSM:
\beq
-{\cal L}_{{\rm SUSY}\>{\rm breaking}} = b H_u H_d
+ {\rm c.c.}
\eeq
Note that $b = h_b \langle X_1 \rangle \langle X_2 \rangle/M_P$ is of
order
$m_W^2$, as required for proper electroweak symmetry breaking. 

The coupling $\epsilon$ and mass parameter $m_{\stilde S}$
parameterize our
ignorance of the high-energy theory. 
The smallness of $\epsilon$ means that $S$ nearly decouples.  
However, the conservation of $R$-parity implies
that if the singlino is the LSP, then decays of the NLSP to
$\stilde S$ will not suffer any competition and can be observed if they
happen within a collider detector.  These decays occur and are
potentially observable because the ``singlino" $\stilde S$ mixes
slightly with the gauginos and Higgsinos, as well as couples directly to
higgsino-Higgs pairs. In that sense, these models are similar to the
well-studied
\cite{GunionHaber}-\cite{Ellwanger:1998vi}
``next-to-minimal supersymmetric standard model" (NMSSM)
\cite{NMSSM}.
The differences include: the extremely small magnitude of $\epsilon$; the
fact that the field $S$ does not obtain a VEV; the absence (or at least
weak-scale phenomenological irrelevance) of an $S^3$ term in the
superpotential; and
the presence of a tree-level supersymmetric mass term for 
$\stilde S$. Nevertheless, it is useful to compare the situation
under study here to a very weakly coupled limit of the NMSSM.
Indeed, the possibility of macroscopic decays involving a
singlino have already been noted in ref.~\cite{Ellwanger:1998vi}, but
considering larger couplings (effectively $\epsilon \gsim 10^{-6}$) and
for smaller values of $m_{\stilde S}$ appropriate for LEP.

If the NLSP is the lightest of the ordinary MSSM neutralinos $\NI$,
then it
can decay according to
\beq
\NI \rightarrow f \overline f \stilde S
\eeq
through virtual sleptons and squarks and virtual or on-shell  $Z$ bosons
and Higgs bosons.
The decay width is 
estimated very roughly by
\beq
\Gamma \sim {m_{\NI} \over 16 \pi} |\epsilon \mu /v|^2 
\times ({\rm suppression}\>{\rm factors}).
\eeq
The suppression factors include the effects of electroweak
couplings, mixing angles, kinematic
suppressions, and (if the mediating boson is not on-shell) three-body
phase space. Without these effects, the rough estimate (for $m_{\NI}
\approx
100$ GeV and $|\epsilon \mu /v| \approx 10^{-8}$) would be of order 
1 meter$^{-1}$ for $\Gamma$. 

After including the suppression effects in realistic models, we will find
that when $\NI$ is allowed to
decay through an on-shell CP-even Higgs boson $h^0$, the inverse decay
width is of order meters or tens of meters. 
Of course, the decay $\NI \rightarrow h^0 \stilde S$ may not be
kinematically allowed. In that case, there may still be allowed
decays $\NI \rightarrow Z^0 \stilde S$. These are typically further
suppressed by a
mixing angle, because the singlino must mix with the MSSM higgsinos in
order to couple to the
$Z$ boson. Nevertheless, we will find that these mixing angles are
typically large enough
so that the inverse decay widths can be of order hundreds of meters.
Finally, it may be that $m_{\NI} - m_{\stilde S} < m_Z$. In that
case, there can still be three-body decays $\NI \rightarrow f\overline
f\stilde S$
through virtual sleptons, squarks, and the $Z^0$. (Decays through
off-shell
Higgs bosons can also occur, but are typically very small because the
MSSM Higgs boson widths are tiny unless they are heavy.) If the $Z$ boson
is far off-shell, then with the usual
model
prejudices that sleptons are much lighter than squarks one finds that
the smallest inverse decay widths are for $\ell^+\ell^-\stilde S$ final
states, and can be of order tens of kilometers. (All
of these
results assume $\epsilon \approx 10^{-8}$, and the decay widths
must be scaled with $\epsilon^2$.)

The majority of decays $\NI \rightarrow
f\overline f \stilde S$
will evidently occur well outside of a typical collider detector. However,
with a
significant number
of supersymmetric events available, a small but finite fraction will
occur inside the detector where the displaced secondary vertex can be
distinguished.
Since the decaying $\NI$ and the resulting $\stilde S$
are invisible, the experimental signature will involve an energetic
lepton-antilepton pair or dijet pair with a significant opening
angle appearing ``from nothing" (with no
corresponding charged particle track pointing back to the interaction
point)
at
a common point. This determination could be accomplished within the inner
tracking volume of a
detector, but might also be possible and perhaps even easier to
distinguish if
the decay occurs farther from the
beam pipe. 
Thus, decays occurring within a meter to several meters
from the interaction point could be a striking, if rare, signal. 

At the Large Hadron Collider, the number of supersymmetric
events expected per year with a low luminosity option of 10
fb$^{-1}$/year can be roughly of order a few thousand to a few million or
more
for 200 GeV $> m_{\NI} >$ 50 GeV. (This assumes $m_{\stilde g} \approx
m_{\stilde q} \approx 7 m_{\stilde N_1}$; of course this is quite
model-dependent.) Every
event gives two possible 
NLSP decays.
Therefore one can aspire to
detect
rare decays with widths as long as hundreds of kilometers by searching
within
the supersymmetric event sample \cite{Hinchliffe:1999ys}. 
In the limit of small decay widths, the probability that a
particular $\NI \rightarrow f \overline f \stilde S$ decay will occur
within a distance $L$ of the interaction point is given by
\beq
P(L) \approx {L \Gamma / \beta \gamma}
\eeq
where $\Gamma$ is the invariant partial width for that decay channel.
Supersymmetric
events will be ``tagged" by the other
particles from the sparticle decays and the presence of large
$\mEt$. 
Since the slowly
decaying $\NI$ will be
massive
and not ultrarelativistic, one can use timing information together with
the pointing information from the tracking detectors and drift chambers 
and perhaps vetoes from the 
muon system to
eradicate backgrounds from cosmic rays and other sources. There have also
been proposals motivated by gauge-mediated supersymmetry breaking
models and by neutralino decays to axino and photon
\cite{Hisano:1997ja} to build
special detector
components
and instrumented tunnels to aid in the
search for very slow decays 
\cite{ChenGunion,Maki:1998ih}. 
At future $e^+e^-$ linear colliders,
supersymmetric event rates are
smaller, but the two-body decays mentioned above might occur often enough
to be detected. The signal is somewhat more problematic at future
runs of the Fermilab Tevatron collider, since the total supersymmetric
event rates are likely to be considerably smaller.
In this paper, I will simply
remain
optimistic
and choose to present results for decay partial widths down to as
small as (1000 km)$^{-1}$.

In these models, there are also singlet scalars $S$ within the
same supermultiplet as $\stilde S$. These very weakly coupled scalars have
masses of order $m_W$ (or, in the case of the invisible axion, essentially
0). However, decays like $\stilde N_1 \stilde S S$ depend on
couplings that are effectively doubly suppressed by $\epsilon$. Other
decays involving only singlet scalars always have competition from
ordinary unsuppressed MSSM decays, and so are not relevant for colliders.

The rest of this paper is organized as follows. In section 2, I 
examine some specific models which realize the idea outlined above.
Section 3 discusses the couplings and mixings of the singlino/axino,
and the relevant decays. Some representative numerical
results for decays to the singlino are presented in section 4.
Section 5 contains some concluding remarks.
An Appendix contains complete formulas for decay widths, including
the effects of arbitrary phases.

\section{Singlino masses and couplings in models with an
intermediate-scale solution to the $\mu$ problem}
\label{sec:models}
\setcounter{equation}{0}
\setcounter{footnote}{1}

Let us consider several models which realize the basic idea outlined
in the introduction.
The magnitude of $f \sim \langle X_{1,2} \rangle$ is (up to
dimensionless couplings) the geometric mean of the Planck scale and
the weak scale in order to agree  with eq.~(\ref{axionwin}). One way
that this could happen is if the soft
supersymmetry-breaking (mass)$^2$ of $X_1$ is driven negative at an
intermediate scale. More
generally, $X_1$ and $X_2$ correspond to a nearly-flat direction in the
potential,
so that dimensionless supersymmetry breaking terms involving $X_1$ and
$X_2$ will
always favor a non-trivial minimum at an intermediate scale. 
The key things we want to show are that these models generically contain
one or more singlet fermions which have electroweak scale (or smaller)
masses, and couplings that are of order $v/f$ (with possible enhancements)
to the MSSM Higgs fields. 

For example, suppose that the superpotential 
contains, in addition to eq.~(\ref{Waxion}), a term 
\beq
W = {\lambda_X \over 6 M_P} X_1 X_2^3,
\label{WXXbarXbarXbar}
\eeq
as in
\cite{Murayama:1992dj}. This fixes the PQ charges of the superfields, 
ensuring the presence of an invisible axion provided that no
other terms break the PQ symmetry.
Here $\lambda_X$ is a dimensionless coupling which is
assumed to be of order unity.
The supersymmetry breaking Lagrangian must then include
\beq
-{\cal L}_{{\rm SUSY} \>{\rm breaking}} = 
m_1^2 |X_1|^2 + m_2^2 |X_2|^2 - {h_X\over 6 M_P}
(X_1 X_2^3 + {\rm c.c.}) 
\label{softbreaking}
\eeq
where $h_X$ is a mass parameter of order the electroweak scale
and has been taken to be real and positive without loss of generality.
A nontrivial global minimum will exist provided e.g.~that
$m_1^2$ is negative at the
scale of the
VEV. However, it is important to note that this is not necessary. The
presence of a holomorphic coupling $h_X$ always favors spontaneous
symmetry breaking at an intermediate scale. So VEVs for $X_1$ and $X_2$
can arise from a negative squared mass, and/or an
$h_X$ which is sufficiently
large \cite{implications,primer}.
In any case, $\langle X_1\rangle$ and $\langle X_2 \rangle$ are of order
$(m_W M_P)^{1/2} \sim
10^{10}$ GeV, which is
naturally within the invisible axion window. 

Let us parameterize the VEVs of $X_1$ and $X_2$ by an overall magnitude
$f$ and an angle $\phi$, so that $\langle X_1 \rangle = f \cos\phi$
and $\langle X_2 \rangle = f \sin\phi$.
Now expand the fields around their VEVs to obtain low-energy
supermultiplet degrees of freedom $S_1, S_2$:
\beq
\pmatrix{ X_1 \cr X_2} =
\pmatrix{\cos\phi \cr \sin \phi} f
+ \pmatrix{
\cos\theta & \sin\theta \cr
-\sin\theta & \cos\theta } \pmatrix{S_1 \cr S_2}.
\label{mixlight}
\eeq
By requiring that the superpotential masses for the fermions $\stilde S_1$
and
$\stilde S_2$ derived from eq.~(\ref{WXXbarXbarXbar})
are diagonal, one can solve for the
mixing angle $\theta$ in terms of the VEV angle $\phi$, with the
result $\theta = \phi/2$. [This typically does not diagonalize
the
axion and other light scalar masses, and depends particularly on the
choice
of eq.~({\ref{WXXbarXbarXbar}).] 
The resulting masses and couplings for $\stilde S_1$ and $\stilde S_2$ are
then found to be
\beq
m_{\stilde S_1} = {\lambda_X f^2\over 2M_P} \sin\phi\, (\cos\phi-1);
\qquad\qquad 
&& \epsilon_{\stilde S_1} = {v \over f}\left (
{\cos\phi/2 \over \cos\phi}-{\sin\phi/2 \over \sin\phi} \right ); 
\label{mepsone}\\
m_{\stilde S_2} = {\lambda_X f^2 \over 2M_P} \sin\phi\, (\cos\phi+1);
\qquad\qquad &&
\epsilon_{\stilde S_2} = {v \over f}\left (
{\sin\phi/2 \over \cos\phi}+{\cos\phi/2 \over \sin\phi} \right ).
\label{mepstwo}
\eeq
[Compare eq.~(\ref{Wlight}).]
The scale $f$ and the angle $\phi$ could also be computed, in principle,
in terms of the
parameters in the soft
supersymmetry-breaking Lagrangian eq.~(\ref{softbreaking}) and the
superpotential. The same parameters also determine the soft
scalar mass of the saxion and other scalars with electroweak scale masses.
However, these will not play any direct role in this paper, so I will not
do this explicitly, and I will treat $f$ and $\phi$ as free parameters. 

One interesting limit is that of small $\phi$ (i.e., $\langle X_2 \rangle$
small compared to $\langle X_1 \rangle$),
in which the mass
eigenstate $\stilde S_1$
is the axino. Then one finds that, up to phases,
\beq
&& m_{\stilde S_1} = {\lambda_X f^2 \over
4M_P}\sin^3\phi;  \qquad\qquad 
\>\, \epsilon_{\stilde S_1} = 
{v \over 2 f};\\
&& m_{\stilde S_2} = {\lambda_X f^2 \over
M_P}\sin\phi;  \qquad\qquad \>\>\>\>
\epsilon_{\stilde S_2} = 
{v \over f\sin\phi}.
\eeq
Note that both masses become small in this limit.
The coupling $\epsilon_{\stilde S_1}$ must grow like $1/\sin\phi$ in
this parameterization
in order for $\mu$ to not become much less than $v$. (LEP has not
discovered a higgsino.)
Another interesting limit is $\phi = \pi/4$ (VEVs of equal magnitude),
resulting in 
\beq
&& m_{\stilde S_1} = 0.10 {\lambda_X f^2 \over
M_P};\qquad\qquad
\>\, \epsilon_{\stilde S_1} = 0.77 v/f;\\
&& m_{\stilde S_2} = 0.60 {\lambda_X f^2 \over
M_P};\qquad\qquad\>\>
 \epsilon_{\stilde S_2} = 1.85 v/f,
\eeq
again, up to phases.
Both $\stilde S_1$ and $\stilde S_2$
have masses that are roughly of order the electroweak scale.
In general they each
contain an
admixture of
the axino. It is interesting to note that $|m_{\stilde S_1}| 
/(\lambda_X f^2/M_P)
< 0.1
$ over the whole range $\theta < \pi/4$. So it is not
unlikely that one or both of $\stilde S_1$ and $\stilde S_2$ is lighter
than all MSSM sparticles.

Other models can be obtained by choosing superpotentials
\beq
W = {\lambda_X \over 6 M_P} X_1 X_2^3
+ {\lambda_\mu \over 2M_P} X_2^2 H_u H_d ,
\label{CCK}
\eeq
as in ref.~\cite{ChoiChunKim}, or 
\beq
W = {\lambda_X \over 6 M_P} X_1 X_2^3
+ {\lambda_\mu \over 2 M_P} X_1^2 H_u H_d .
\label{notCCK}
\eeq
In both of these cases, the diagonalized singlino masses are still given
as in eqs.~(\ref{mepsone}) and (\ref{mepstwo}), since they only depend on
the
$\lambda_X$ term in the superpotential. However, the couplings
are modified to, respectively:
and
\beq
\epsilon_{\stilde S_1} = -{2 v \sin\phi/2\over f \sin\phi};\qquad
\epsilon_{\stilde S_2} = {2 v \cos\phi/2\over f \sin\phi}.
\eeq
for eq.~(\ref{CCK}), and 
\beq
\epsilon_{\stilde S_1} = {2 v \cos\phi/2\over f \cos\phi};\qquad
\epsilon_{\stilde S_2} = {2 v \sin\phi/2\over f \cos\phi}
\eeq
for eq.~(\ref{notCCK}).

Another similar model is obtained by assuming a different form of the
$\lambda_X$ term used to stabilize the potential at large field strengths:
\beq
W =  
{\lambda_X \over 4 M_P} X_1^2 X_2^2 
+{\lambda_\mu \over 2 M_P} X_1^2 H_u H_d .
\label{XXXbarXbar}
\eeq
Again using the mixing parameterization
eq.~(\ref{mixlight}),
one finds in this class of models that now $\theta = (1/2) \tan^{-1}(2
\tan 2\phi)$
in order that $\stilde S_1$ and $\stilde S_2$ are mass eigenstates.
In terms of the VEV angle $\phi$, we find:
\beq
m_{\stilde S_1} & =& {\lambda_X f^2 \over 4M_P} 
\left [ 1 - \cos 2\phi \sqrt{1 + 4 \tan^22\phi} \right ];
\\
\epsilon_{\stilde S_1} &=& {\sqrt{2} v \over f \cos\phi}
\left [ 1 +  {1\over\sqrt{ 1 + 4 \tan^22\phi}} \right ]^{1/2};
\\
m_{\stilde S_2} &=& {\lambda_X f^2 \over 4M_P} 
\left [ 1 + \cos 2\phi \sqrt{1 + 4 \tan^22\phi} \right ];
\\
\epsilon_{\stilde S_2} &=& {\sqrt{2} v \over f \cos\phi}
\left [ 1 -  {1\over\sqrt{ 1 + 4 \tan^22\phi}} \right ]^{1/2},
\eeq
up to phases. In the limit $\phi \rightarrow 0$, one finds
\beq
&& m_{\stilde S_1} = {3 \lambda_X f^2 \over 2 M_P}\sin^2\phi;
\qquad\qquad
\epsilon_{\stilde S_1} = {2 v\over f};\\
&& m_{\stilde S_2} = { \lambda_X f^2 \over 2 M_P};
\qquad\qquad\qquad\>\>\>
\epsilon_{\stilde S_2} = 
{4v\over f}
\sin\phi ,
\eeq
which features a parametric suppression in the mass of $\stilde S_1$. In
the case of equal VEVs $\phi = \pi/4$, one obtains instead
\beq
&& m_{\stilde S_1} = {\lambda_X f^2 \over 4 M_P};\qquad\qquad
\>\,\epsilon_{\stilde S_1} = {2 v\over f};\\
&& m_{\stilde S_2} = {3\lambda_X f^2 \over 4 M_P};\qquad\qquad
\epsilon_{\stilde S_2} = {2 v\over f}.
\eeq
Finally, the limit $\phi \rightarrow \pi/2$ 
(i.e., $\langle X_1 \rangle$ small compared to $\langle X_2 \rangle$)
yields a
very light
$\stilde
S_2$ and a relative enhancement of the $\stilde S_1$ coupling:
\beq
&& m_{\stilde S_1} = {\lambda_X f^2 \over 2 M_P};\qquad\qquad\qquad
\>\>\>\> \epsilon_{\stilde S_1} = {2 v\over f \cos\phi };\\
&& m_{\stilde S_2} = {3\lambda_X f^2 \over 2 M_P}\cos^2\phi;\qquad\qquad
\epsilon_{\stilde S_2} = {4 v\over f}.
\eeq

There are clearly many possible more complicated variations on these
models; for example,
schemes with more than two fields $X_i$ participating in the PQ-breaking
and $\mu$-generating dynamics (see, for example, 
ref.~\cite{Cleaver:1998nj}). It
is also possible to
have a
scheme in which there is only one field $X$, which obtains a VEV
at an intermediate scale below where the soft mass term $m_X^2$ runs
negative.
This corresponds to the $\phi \rightarrow 0, \lambda_X \rightarrow 0$
limit of
eq.~(\ref{XXXbarXbar}) with $\stilde S_2$ removed, so there is just a
light
axino with $\epsilon = 2v/f$.
The essential features of all these models are that they contain
one or more singlino
fields, with couplings naively of order $v/f$ but which can be
significantly enhanced,
and which can easily be lighter than
all of the MSSM odd-$R$-parity sparticles. 
 
\section{Mixing of the singlino with MSSM neutralinos}
\label{sec:mixings}
\setcounter{equation}{0}
\setcounter{footnote}{1}

As shown in the previous section, one or both singlino mass eigenstates
$\stilde S_1$ or $\stilde S_2$ can be lighter than all MSSM sparticles.
In this section, I will consider the relevant mixings and couplings of
such a singlino to the MSSM states, and the ensuing decay partial widths.
I will use $\stilde S$ to refer generically to either $\stilde S_1$ or
$\stilde S_2$.

The properties of the singlino are determined by the superpotential
eq.~(\ref{Wlight}). At tree level, there are singlino-higgsino-Higgs boson
couplings. Other couplings, including singlino-fermion-sfermion and
singlino-higgsino-$Z$ boson, arise due to singlino-gaugino and
singlino-higgsino mixing.
In order to discover the
couplings of the $\stilde S$ to the physical MSSM states, one must
diagonalize the
the $5\times 5$
neutralino mass matrix. In the 
($\stilde S$, $\stilde B$, $\stilde W^0$, $\stilde H_d^0$, 
$\stilde H_u^0$) basis, it is given by:
\beq
M^{(5)} = 
\pmatrix{ 
m_{\stilde S} & 0 & 0 & -\epsilon\mu\sbeta & -\epsilon\mu\cbeta \cr
0 & M_1 & 0 & - \cbeta\, \sW\, \mZ & \sbeta\, \sW \, \mZ  \cr
0 & 0 & M_2 & \cbeta\, \cW\, \mZ & - \sbeta\, \cW\, \mZ \cr
-\epsilon\mu\sbeta & -\cbeta\,\sW\,\mZ & \cbeta\,\cW\,\mZ & 0 & -\mu\cr
-\epsilon\mu\cbeta & \sbeta\,\sW\,\mZ & -\sbeta\,\cW\,\mZ & -\mu & 0\cr 
},
\label{Mfive}
\eeq
where $s_\beta, c_\beta$ stand for $\sin\beta$, $\cos\beta$, and $s_W,c_W$
for $\sin\theta_W$, $\cos\theta_W$.
In the following, I shall take $m_{\stilde S}$ to be real and positive
without loss of generality. This allows $\epsilon$ to have an
arbitrary phase. Now, the off-diagonal terms proportional to
$\epsilon$ can be treated as a perturbation. 
Therefore, our procedure
is to first diagonalize $M^{(4)}$, the lower right $4\times 4$ mass
sub-matrix. This is accomplished
with a unitary matrix $Z_{ij}$ $(i,j=1,\ldots,4)$ according to:
\beq
Z_{ik}^* M^{(4)}_{kl} Z^{*}_{jl} = \delta_{ij} m_{\stilde N_j}.
\label{diagfour}
\eeq
Here the masses $m_{\stilde N_j}$ are real and positive; this can always
be done, regardless of the relative complex phases of $\mu$, $M_1$ and
$M_2$. 
To the lowest order in a perturbative expansion in $\epsilon$,
the singlino $\stilde S$ is a mass eigenstate, and the ordinary MSSM
neutralinos
have the same masses that they would have had if $\stilde S$ were absent.
I will choose an ordering scheme
such that $\stilde S = \stilde N_0$, with $m_{\stilde S} 
= m_{\stilde N_0} < m_{\stilde N_1} < m_{\stilde N_3} < m_{\stilde N_3}<
m_{\stilde N_4}$. 

The full $5 \times 5$ neutralino-singlino mass matrix
can then be
diagonalized according to
\beq
N_{ik}^* M^{(5)}_{kl} N^{*}_{jl} = \delta_{ij} m_{\stilde N_j} ,
\label{diagfive}
\eeq
where now $i,j = 0,1,\ldots,4$. 
To lowest order in a perturbation in $\epsilon$, one 
finds that $N_{00} = 1$ and $N_{ij} = Z_{ij}$ for $i,j = 1,2,3,4$.
So one can write
\beq
N_{ij} = \pmatrix{ 1 & N_{0j}\cr N_{i0} & Z_{ij} } .
\label{Zfive}\eeq
The neutralino-singlino
mixing elements can be determined in terms of the $4\times 4$ $Z_{ij}$,
the mass eigenvalues, and the parameter $\epsilon$:
\beq
N_{0j} &=& \sum_{k=1}^4 Z_{kj}
{\left [\epsilon^* \mu^* m_{\stilde N_k}
(\sbeta Z_{k3} + \cbeta Z_{k4}) + \epsilon \mu m_{\stilde S}
(\sbeta Z^*_{k3} + \cbeta Z^*_{k4})\right ] 
/
(m_{\stilde N_k}^2 - m_{\stilde S}^2)
};\phantom{xxx}
\label{Nfivei}\\
N_{i0} &=& 
-\left [
\epsilon \mu m_{\stilde N_i} 
(\sbeta Z^*_{i3} + \cbeta Z^*_{i4})
+ \epsilon^* \mu^* m_{\stilde S}
(\sbeta Z_{i3} + \cbeta Z_{i4})
\right ]/
(m_{\stilde N_i}^2 - m_{\stilde S}^2) .
\label{Nifive}
\eeq
(Again, nothing has been assumed here about the complex phases 
of the parameters $M_1$, $M_2$, $\mu$ and $\epsilon$.) 
This expansion is an excellent approximation because $|\epsilon|$ is
very small. In particular, one can check by numerical
diagonalization of eq.~(\ref{Mfive}) that the perturbative result is very
accurate unless the denominator $m_{\stilde N_1}^2 - m_{\stilde S}^2$ in
eqs.~(\ref{Nfivei}) and
(\ref{Nifive}) is tuned to 0 to
an extreme accuracy (comparable to $|\epsilon \mu m_{\stilde S}|$),
in
which case the decay widths studied below are completely
negligible anyway. 
 
The relevant neutralino and singlino couplings for low energy
phenomenology are
all contained in
the five eigenmasses $m_{\stilde N_i}$ and $m_{\stilde S}$ (for which
corrections proportional to $\epsilon$ are negligible) and the neutralino
mixing quantities $N_{ij}$ and $N_{0j}$ with ($i,j=1,\ldots,4$).
The decay rates with a singlino in the final state can now be worked
out, as in the NMSSM \cite{Franke}. I present
formulas for the relevant widths in an Appendix, taking care to allow for
possible arbitrary phases in the neutralino mixing matrix and in
$\epsilon$. 
In general, the phase of $\epsilon$ could be anything, since it is
not constrained by low-energy experiments on CP violation and is
not necessarily correlated with the phase of $\mu$.
Because the results might be useful for other problems, I will provide
the general results for a decay involving $\stilde N_i$
to $\stilde N_j$; the case needed in this paper is obtained by
simply taking
$i =1$ and $j=0$.
Numerical illustrations of these results will be given in Section 4.

The necessary
coupling for the decay
\beq 
\stilde N_1 \rightarrow h^0 \stilde S
\eeq 
arises directly from the superpotential eq.~(\ref{Wlight}) through
the
higgsino content of $\stilde N_1$. It also obtains contributions from
the singlino-higgsino mixings $N_{03}$ and $N_{04}$
combined with the gaugino
content of $\stilde N_1$, and from the singlino-gaugino mixing elements
$N_{01}$
and $N_{02}$ combined with the higgsino
content of $\stilde N_1$. The total coupling is given explicitly
by eq.~(\ref{GNSh}). The Higgs will then decay according to $h^0
\rightarrow b\overline b$, $WW^*$, $\tau^+\tau^-$, 
$c \overline c$,
or $g \overline g$, providing a visible product displaced from the
original interaction point producing the event.

The decay
\beq
\stilde N_1 \rightarrow Z \stilde S 
\eeq
relies both on singlino-higgsino mixings $N_{03}$ and $N_{04}$
and on
the higgsino
content of $\stilde N_1$, and is therefore somewhat more suppressed in
models with a bino-like NLSP. The relevant coupling is given explicitly
by eq.~(\ref{GNSZ}). The $Z$ boson then decays with Standard Model
branching fractions to quark-antiquark and lepton-antilepton pairs.

When neither two-body decay is open, the neutralino NLSP will decay
through off-shell sleptons, squarks, the $Z$, and Higgs bosons. The
full expressions for these decays are given in the Appendix. The most
important contribution for a bino-like NLSP is typically through sleptons,
and therefore relies on the singlino-bino and singlino-wino mixing
elements $N_{01}$ and $N_{02}$. Fortunately,
these are not greatly suppressed in many models unless $|\mu|$ is very
large.

There remains the possibility
of a two-body decay $\NI \rightarrow \gamma \stilde S$, which arises at
the one-loop level. However, using the results of
\cite{Haber:1989px} (with appropriate changes for the couplings
to correspond to the model under present consideration), I have verified
that these decays are always suppressed compared to the
two- and three-body decays considered here, with widths that cannot be 
much larger than a few times (1000 km)$^{-1}$ in the examples 
in the next section with $\epsilon = 10^{-8}$. Since these decays are
not competitive here, I will
not present results for them.

A separate possibility is that the NLSP is a stau, or that all three
lighter, mostly-right-handed, slepton mass eigenstates ($\stilde \tau_1,
\stilde \mu_R, \stilde e_R$) have no open decays except to the singlino.
In that case, one can hope to observe $\stilde \tau_1 \rightarrow \tau
\stilde S$ (and perhaps $\stilde \mu_R \rightarrow \mu \stilde S$ and
$\stilde e_R \rightarrow e \stilde S$). The decaying slepton will appear
in the detector as a muon-like charged particle track, or as a track
with an anomalously high ionization rate. The rare decay to the singlino
will yield a large-angle kink in the track leading either to
a tau jet or an electron or muon. Since the decaying particle is
heavy, there will be a significant angle at the kink. The decay widths are
suppressed only by the singlino-gaugino mixing, so they can occur
within the detector often enough to measure, even if
$\epsilon$ is significantly less than $10^{-8}$.

\section{Representative results for decays to the singlino}
\label{sec:results}
\setcounter{equation}{0}
\setcounter{footnote}{1}

In this section, I will consider some illustrative numerical results for
decays to the singlino, first for neutralino NLSP models and then for stau
or slepton NLSP models.  I will take $\epsilon = 10^{-8}$, 
with the understanding that the results have to be scaled
according to $\Gamma \propto \epsilon^2$.

\subsection{Neutralino decays}

In order to study the decay partial widths of a neutralino, I will
employ the concept of ``model lines", in which one supersymmetry-breaking
parameter is allowed to vary, setting the overall scale for all sparticle
masses. 
First, consider
a typical model scenario with a bino-like NLSP and the LSP singlino
mass fixed at $m_{\stilde S} = 50$ GeV. 
The bino mass parameter
$M_1$ is varied, with the wino mass parameter $M_2$ and the
$\mu$ term then determined according to $M_2 = 2.0 M_1$; $\mu = 3.0 M_1$. 
The $5\times 5$ neutralino mass matrix is then fully determined by also
choosing
fixed values of $\tan\beta = 3.0$ and $\epsilon = 10^{-8}$.
The right-handed slepton masses $m_{\stilde e_R}=m_{\stilde
\mu_R}=m_{\stilde \tau_R}$
are constrained to be the greater of 
$1.2 m_{\stilde N_1}$ and 110 GeV. This assures that a slepton
cannot be the NLSP and should
not be found at LEP. Mixing in the stau sector is
neglected. The left-handed slepton masses are determined by $m_{\stilde
e_L}^2 = m_{\stilde e_R}^2 + 0.5 M_2^2$. I will assume that squarks are
not light enough to give a significant
contribution to the decay. Finally, the lightest Higgs boson
mass is assumed to be $m_{h^0} = 120$ GeV, safely out of the reach
of LEP, and to obey the decoupling limit $\alpha = \beta - \pi/2$.
\begin{figure}[tbp] 
\centering\epsfxsize=4.2in 
\hspace*{0in}
\epsffile{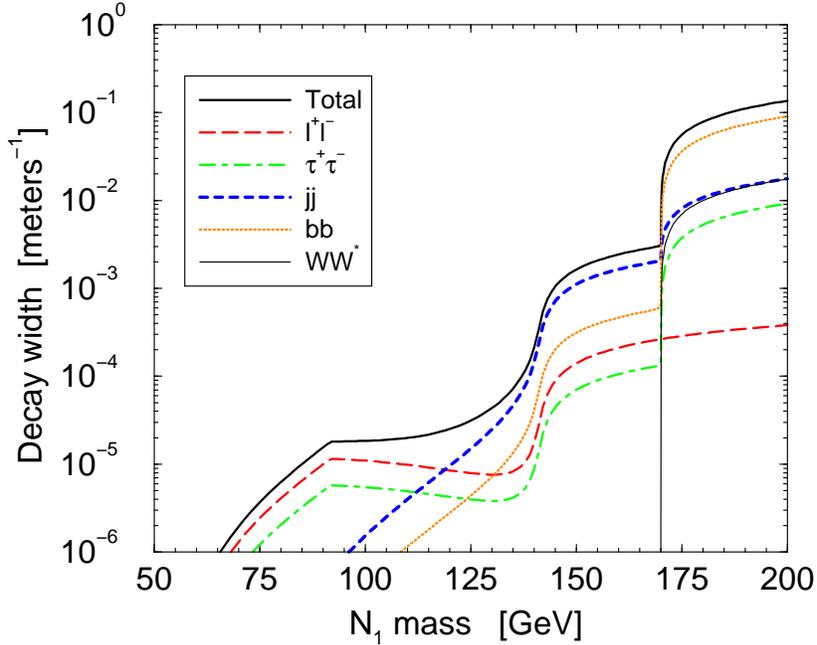}
\caption{Visible decay widths of bino-like neutralinos $\stilde N_1$ into
final states involving the singlino, as a function of varying
$m_{\stilde N_1}$ for fixed $m_{\stilde S} = 50$ GeV and $\epsilon =
10^{-8}$. 
The MSSM model parameters 
are described in the text.
The solid line is the total visible decay width. The long dashed line is
the partial width into $\ell^+\ell^-\stilde S$ where $\ell = e$ or $\mu$.
The dot-dashed line is the width into $\tau^+\tau^-\stilde S$.
The short-dashed line includes $jj \stilde S$ where $j$ is any 
$u,d,s,c$ (anti)-quark jet or gluon jet, and the dotted line is for
$b \overline b \stilde
S$. The thin solid line is the width for $W W^*\stilde S$ through an
on-shell Higgs boson.
}
\label{width50}
\end{figure}
The results for the partial decay widths to visible states (excluding
neutrinos), as found from the equations in the Appendix, are shown
in Figure \ref{width50} as a function of $m_{\stilde N_1}$. 

For $m_{\stilde N_1} \lsim 120$ GeV in this model line, the decays are
dominated by the contributions of the virtual right-handed sleptons. The
total inverse decay lengths are of order tens of kilometers, and are
nearly 
democratic between $e^+e^-$, $\mu^+\mu^-$ and $\tau^+\tau^-$ final states.
The ``knee" near $m_{\stilde N_1} = 92$ GeV is merely an artifact of the
constraint $m_{\stilde e_R} > 110$ GeV; for smaller neutralino masses, the
virtual slepton is necessarily more off-shell due to the constraint that
it has not been discovered at LEP.

For $m_{\stilde N_1} \gsim 120$ GeV, the contributions from the virtual
$Z$ boson [the terms $W_Z$, $W_{Zt}$, and $W_{Zu}$ in
eq.~(\ref{threebody})] begin to be important. This increases the decay
partial widths, with contributions to $f\overline f$ that are
roughly proportional to the $Z$ branching fraction, so that dijet final
states dominate. For $m_{\stilde N_1} > 142$ GeV, the virtual Z boson is
on-shell, and the decay becomes two-body $\NI \rightarrow Z^0 \stilde S$.
(The three-body formula is used in the vicinity of threshold, however,
in order to correctly include interference effects with the virtual
slepton diagrams in that regime.)
This leads to a total visible
decay width greater than (1000 meters)$^{-1}$. For negative $\mu$, the
decay widths tend to be somewhat smaller.

Finally, for $m_{\stilde N_1} > m_{\stilde S} + m_{h^0} = 170$ GeV in this
model line, the decay $\NI \rightarrow h^0
\stilde S$ opens up and completely dominates. Since there is a direct
higgsino-singlino-Higgs boson coupling, this is much larger than
the two-body decay to $Z\stilde S$, even though the $Z$ boson is lighter.
The results are shown assuming Standard Model branching fractions for
$h^0$ into final states $b\overline b$, $WW^*$, $\tau^+ \tau^-$, and
(lumped together into the ``$jj$" category)
$c\overline c$ and $gg$.
Here the partial decay width of $\NI$ to the $b \overline b
\stilde S$
final state is found to be of order (10 meters)$^{-1}$. Of course, if the
$h^0$ mass is smaller, this mode will open up and dominate for smaller
values of $m_{\NI}$.
 
In the era after supersymmetry is discovered, the situation will be
rather different; we will presumably know the MSSM sparticle
mass spectrum, but the singlino mass and coupling will be completely
unknown. So, a more useful summary of the situation we could face might be
something like
that shown in Figure \ref{widths}. This is a particular point 
along the same model
line just discussed, with $m_{\stilde N_1} = 150$ GeV, and with the
horizontal axis
representing the possible values of the singlino mass.
\begin{figure}[tbp] 
\centering\epsfxsize=4.2in\hspace*{0in}
\epsffile{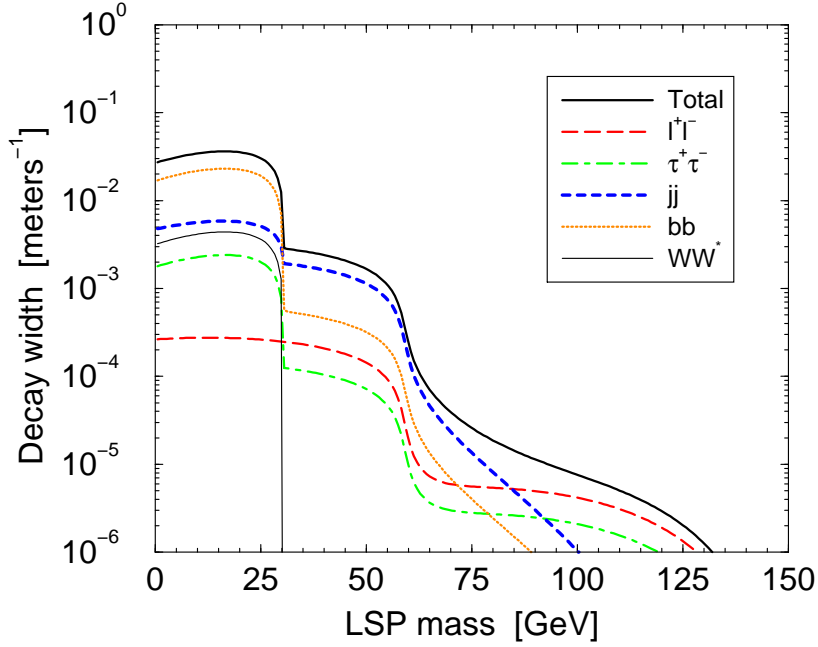}
\caption{As in Figure 1, but for fixed $m_{\stilde N_1} = 150$ GeV,
and varying $\stilde S$ (LSP) mass.} 
\label{widths}
\end{figure}

A somewhat different scenario ensues if the NLSP is a higgsino-like
neutralino. To illustrate this, I choose a pair of model lines with
$\mu = \pm 0.8 M_1$, and all other parameter relationships as described
above for Figure \ref{width50}. The results are shown in Figure
\ref{widthhiggsino}, but now only for the total visible decay width. Since
$\stilde N_1$ has a smaller gaugino
content, the decays through virtual sleptons are highly suppressed.
Conversely, the large higgsino component of $\NI$ enhances the
probability of decay through a virtual $Z$ boson. 
So, for $m_{\stilde N_1} < m_{\stilde S} + m_{\stilde h^0} = 170$ GeV
in this model line, the $f \overline f$ decays will obey $Z$ boson
branching fractions.
However,
there turns out to be an accidental
suppression of the matrix element for $\mu < 0$ in the convention 
specified in eq.~(\ref{Mfive}) (which is the same as in
refs.~\cite{HaberKane,primer,GunionHaber}),
particularly
when the $Z$ boson is off-shell. For $m_{\stilde N_1} > 170 $ GeV, the
decay $\stilde N_1 \rightarrow h^0 \stilde S$ length
is of order several meters if $\stilde N_1$ is mostly higgsino. 
\begin{figure}[tbp] 
\centering\epsfxsize=4.1in\hspace*{0in}
\epsffile{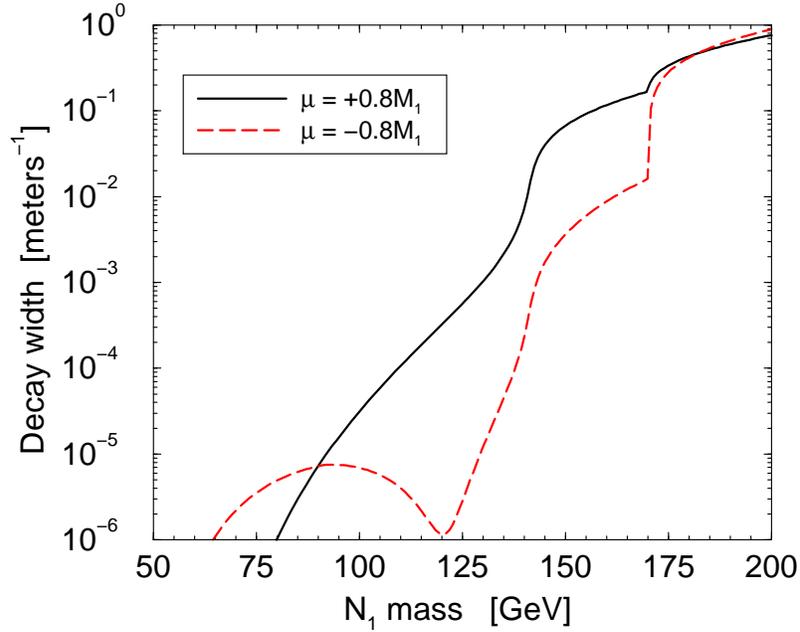}
\caption{Total visible decay widths of higgsino-like
neutralinos $\stilde N_1$  into final states $f \overline f \stilde S$, as
a function of varying $m_{\stilde N_1}$ for fixed $m_{\stilde S} = 50$
GeV. The model parameters satisfy the constraints $M_2 =
2.0 M_1$, $\tan\beta = 3.0$, and $\mu = +0.8 M_1$ (solid line) or $\mu =
-0.8 M_1$ (dashed
line), and other constraints described in the text.} 
\label{widthhiggsino}
\end{figure}

In the above analyses, I have assumed that $H^0$ and $A^0$ are very heavy,
and that three-body amplitudes involving them are negligible. Although
this is appropriate throughout most of parameter space, it is possible
that for large $\tan\beta$, the couplings of $H^0$ and $A^0$ could be
large enough to make an appreciable contribution to $\Gamma(\NI
\rightarrow b\overline b \stilde S)$, even if $H^0$ and $A^0$ are far
off-shell.

Finally, I note that the decay widths discussed here depend on the
phase of the parameter $\epsilon$. This phase is not constrained by
low-energy CP violating observables, and so might be considered
completely arbitrary. I have checked, using the formulas in the Appendix, 
that varying Arg($\epsilon$), while keeping all other
parameters fixed, can change the $\stilde N_1$ decay widths by an order of
magnitude or
so. In models with a bino-like NLSP as in Figure 1, the largest decay
widths tend to occur for real $\epsilon$.

\subsection{Slepton decays}

It is also possible that the NLSP is a slepton $\stilde \tau_1$ or,
effectively,
all three mainly right-handed sleptons $\stilde e_R$, $\stilde \mu_R$
and $\stilde \tau_1$. The latter scenario is realized if $\tan\beta$
is not too large, so that the three sleptons are mass-degenerate
to within less than $m_\tau$. These possibilities are familiar 
\cite{scottandfriends}-\cite{BMTW}
in gauge-mediated supersymmetry breaking models \cite{GMSB}
but could also be realized
in supergravity-mediated models if there is not a large universal
contribution to scalar masses, or if $D$-term contributions
(proportional to some exotic $U(1)$ quantum number) are large.

If $\stilde \tau_1$ is the NLSP, then the two-body decays $\stilde \tau_1
\rightarrow \tau\stilde S$ are suppressed only by the bino-singlino
mixing. In Figure \ref{widthstau}, I show the results for this decay width
for a typical model line with varying $M_1$ and fixed LSP mass $m_{\stilde
S}$. In order to ensure that a stau is the NLSP, the constraint
$m_{\stilde \tau_1} = 0.9 m_{\NI}$ (thick lines) and $0.7 m_{\NI}$
(thin lines) are
imposed. Other
relevant model
line parameters are $ M_2 = 2.0 M_1$, $\mu = 3.0 M_1$, $\tan\beta = 3.0$.
To a good approximation, the decay width depends only on 
the absolute
value of 
$s_{\stilde \tau}$ in the stau mixing parameterization of
eq.~(\ref{smixing}), so results are shown for
$s_{\stilde \tau} = 0$, $0.25$, and $0.5$.
The solid line is also approximately true for $\stilde e_R \rightarrow e
\stilde S$ and $\stilde \mu_R \rightarrow \mu \stilde S$ as a function of
$m_{\stilde e_R}$ and $m_{\stilde \mu_R}$,
by taking $s_{\stilde \tau} = 0$, $c_{\stilde \tau} =1$. 
\begin{figure}[tb] \centering\epsfxsize=4.1in\hspace*{0in}
\epsffile{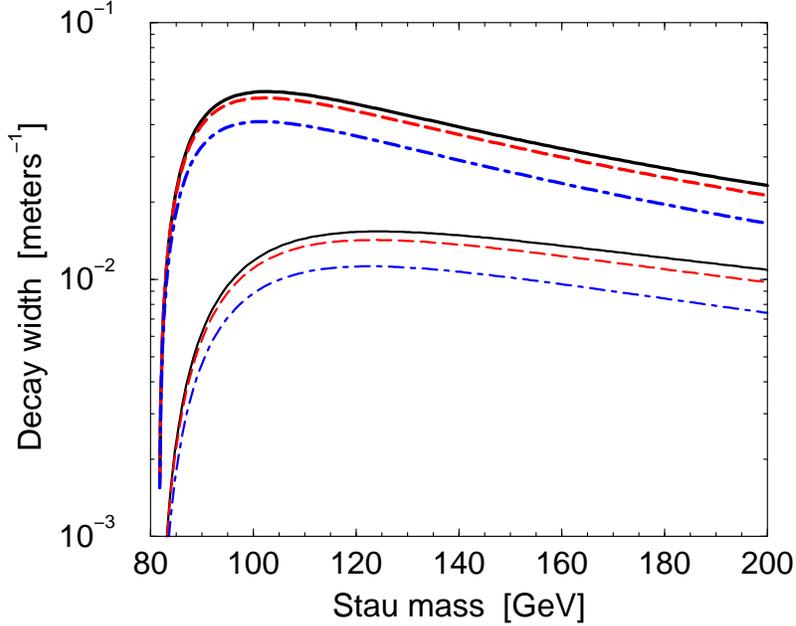}
\caption{Decay widths for $\stilde \tau_1 \rightarrow \tau \stilde S$ as a
function of varying $m_{\stilde \tau_1}$. The model parameters satisfy
$m_{\stilde S} = 80$ GeV, and the constraints $M_2 = 2.0 M_1$, 
$\mu = 3.0 M_1$, and $\tan\beta = 3.0$.
Solid lines are for the unmixed case $s_{\stilde \tau} = 0$, the dashed
lines
for $s_{\stilde \tau} = 0.25$ and dash-dotted lines for
$s_{\stilde \tau} = 0.5$. The thicker (thinner) lines are for
$m_{\stilde \tau_1}/m_{\stilde N_1} = 0.9$ (0.7).} 
\label{widthstau}
\end{figure}

Because there is relatively little suppression in this case, the
inverse decay widths are of order tens of meters. This is
discernible at a collider which can produce several hundred supersymmetric
events. Each such event would contain a pair of quasi-stable stau or
slepton highly ionizing tracks, which can have an anomalously high $dE/dx$
to distinguish them from muons. In a small fraction of events,
one of the stau or slepton tracks will have a kink leading to a lepton
or tau jet, corresponding to
the decay. The resulting tau or lepton would have a significant
angle with respect to the original highly ionizing track, yielding
a potentially spectacular and nearly background-free signal.

\section{Conclusions}
\label{sec:conclusions}
\setcounter{equation}{0}
\setcounter{footnote}{1}  

The presence of the $\mu$ term in the MSSM and the solution
to the strong CP problem may have a common 
explanation at an intermediate scale. Direct detection of the 
resulting axion
is quite problematic. In this paper, I have argued that these 
models may nevertheless give rise to observable signals at colliders,
through delayed decays to singlino fermions that include the axino as
a mixture. These events will be a rare (perhaps very rare) occurrence
within a large sample of supersymmetric events at the Large Hadron
Collider or a future $e^+e^-$ linear collider. 
There is also a possibility that the lightest MSSM
sparticle could have slow decays into more than one singlino.
Note that the LHC cross sections can be very large precisely when the NLSP
decay widths are small.

The numerical estimate in section 4 of this paper have used 
$\epsilon = 10^{-8}$ for the singlino-higgsino-Higgs coupling parameter.
Of
course, the actual value could be significantly smaller. On the other
hand, I showed that
in some models the couplings are parametrically enhanced, and the
mass of one or more singlinos is reduced, 
if one of
the VEVs giving rise to the $\mu$ term is relatively small. Furthermore,
the magnitude of $\epsilon$ can be significantly larger if the high scale
$M_P$ is replaced by a somewhat lower scale that governs
non-renormalizable operators, for example a string scale or a
compactification scale that is not far above the apparent gauge coupling
unification scale. 

Future planning and analysis of collider physics experiments should
take into account the possibility that the apparent LSP is actually
unstable. Besides the models I have discussed here, there are at least
two other plausible variations on the MSSM which can lead to delayed
rare decays of what might appear, at first, to be the stable LSP. 

First, gauge-mediated supersymmetry-breaking (GMSB) models \cite{GMSB}
with a
supersymmetry-breaking scale $\sqrt{F}$ that is not too large will
give rise to decays that could have macroscopic proper lengths
\cite{scottandfriends}-\cite{BMTW}. It is interesting to
compare the reason for this to
that in the models discussed in the present paper. In GMSB models, an
estimate
for a decay width of the NLSP to the goldstino/gravitino $\stilde G$ is
$\Gamma \sim m_W^5/16\pi (\sqrt{F})^4$, while in the decays to a
singlino/axino
LSP $\stilde S$, the
estimate is $\Gamma \sim m_W^3/16 \pi f^2$. 
So NLSP decays are suppressed by the 4th power of the
supersymmetry-breaking scale in GMSB, but only by the square of the PQ
scale in light axino/singlino models. GMSB models can, in fact, give rise
to signals which might
be very difficult to distinguish from those discussed here. For example,
if the NLSP is a neutralino with a significant higgsino content,
it can have \cite{scottandfriends,akkmm,bmpz,HZGMSB} decays $\stilde N_1
\rightarrow h^0 \stilde G$
and
$\stilde N_1 \rightarrow Z^0 \stilde G$ that look like
the decays discussed in this paper. Or, if a stau is the NLSP, it can 
appear quasi-stable with rare decays $\stilde \tau_1 \rightarrow
\tau \stilde G$ occurring within the detector. 
Second, one can have weak $R$-parity violating couplings
\cite{rparityviolation} in the MSSM which could also give decays like
$\stilde N_1 \rightarrow \ell^+ \ell^{\prime -} \nu$ or $\stilde N_1
\rightarrow q \overline q^\prime \nu$. These signatures could mimic
those discussed in the present paper.

If these signals appear, it will be interesting to try to establish
the correct explanation from among the competing hypotheses. The prize for
doing so will be that we will gain an understanding of physics at
scales far above those probed by direct sparticle production at colliders.

\smallskip \noindent {\it Acknowledgements: } I thank Howard Haber,
Janusz Rosiek, and James Wells 
for helpful discussions. This work was supported in part
by National
Science Foundation grant number PHY-9970691.

\addcontentsline{toc}{section}{Appendix: Complete formulas for 
neutralino decay widths including the effects of 
arbitrary phases}\section*{Appendix: Complete formulas for 
neutralino decay widths including the effects of arbitrary phases}
\label{sec:appendix}
\renewcommand{\theequation}{A.\arabic{equation}} 
\setcounter{footnote}{1}

In this appendix, I give formulas for the two- and three-body decays of
a neutralino to another neutralino and a Higgs boson, $Z$ boson, or
fermion-antifermion pair. The model is the extension of the MSSM with one
singlet superfield as specified by eq.~(\ref{Wlight}). This also includes
the
MSSM and NMSSM as special cases. The results needed in Section 4 of this
paper are obtained by taking the neutralino mass eigenstate indices to
be $i = 1$ and $j=0$ in the following.

First, let us consider the two-body decay of a neutralino to another
neutralino and the lightest CP-even neutral Higgs boson. The
relevant decay width is equal to:
\bea
\Gamma(\stilde N_i \rightarrow h^0 \stilde N_j) & = &
{m_{\stilde N_i} \over 16 \pi} \sqrt{\lambda( 1, {
r_{j}^2}, {r_{h^0 }^2} )}
\Bigl (
|G^{h^0}_{ij}|^2 (1 +
{r_{j}^2}- 
{r_{h^0}^2}) 
+ 2 {\rm Re}[(G^{h^0}_{ij})^2] r_j
\Bigr ){\phantom{xxxx}}
\eea
where $r_{j} = m_{\stilde N_j}/m_{\stilde N_i}$; 
$r_{h^0} = m_{h^0}/m_{\stilde N_i}$; $\lambda(a,b,c) = 
a^2 + b^2 + c^2 - 2 a b - 2 a c - 2 b c$; and the
neutralino-neutralino-Higgs
coupling is given by
\beq
G^{h^0}_{ij} &=& 
{1\over 2} (g N^*_{i2} - g' N^*_{i1})
(s_\alpha N^*_{j3}+ c_\alpha N^*_{j4} ) 
+{\epsilon \mu \over \sqrt{2} v}(c_\alpha N^*_{i3} - s_\alpha
N^*_{i4})N^*_{j0}
+ (i \leftrightarrow j).
\phantom{xx}\label{GNSh} 
\eeq
Here 
$c_\alpha$ and $s_\alpha$
denote $\cos\alpha$ and $\sin\alpha$
with $\alpha$ the Higgs mixing angle in the notation of
\cite{GunionHaber}.

The results for the decays to the heavier
neutral CP-even ($H^0$) and CP-odd ($A^0$) Higgs bosons can be obtained by
instead using the couplings:
\beq
G^{H^0}_{ij} &=& 
{1\over 2} (g N^*_{i2} - g' N^*_{i1})
(-c_\alpha N^*_{j3}+ s_\alpha N^*_{j4} ) 
+{\epsilon \mu \over \sqrt{2} v}(s_\alpha N^*_{i3} + c_\alpha
N^*_{i4})N^*_{j0}
+ (i \leftrightarrow j)\phantom{xxx}
\label{GNSH} 
\eeq
for $\stilde N_i \rightarrow H^0 \stilde N_j$, and
\beq
G^{A^0}_{ij} &=& 
{i\over 2} (g N^*_{i2} - g' N^*_{i1})
(s_\beta N^*_{j3}- c_\beta N^*_{j4} ) 
+ i {\epsilon \mu \over \sqrt{2} v}(c_\beta N^*_{i3} + s_\beta
N^*_{i4})N^*_{j0}
+ (i \leftrightarrow j)\phantom{xxx}
\label{GNSA} 
\eeq
for $\stilde N_i \rightarrow A^0 \stilde N_j$, and
substituting $m_{h^0} \rightarrow$ $m_{H^0}$ or $m_{A^0}$ in the obvious
way. However, in the 
numerical
analyses of Section 4, $H^0$ and $A^0$ are assumed to be
heavy and decoupled, so these decays are neglected.
 
Similarly, the two-body decay of a neutralino to another neutralino
and a $Z$ boson has a width given by 
\bea
\Gamma(\stilde N_i \rightarrow Z^0 \stilde N_j) 
= 
{m_{\stilde N_i} \over 16 \pi} \sqrt{\lambda( 1, {
r_{j}^2}, {r_{Z }^2} )}
\Bigl (
|G^{Z}_{ij}|^2 \bigl [1 +
{r_{j}^2}- 2 r_Z^2 + {(1 - r_{j}^2)^2/r_Z^2} \bigr ]
+ {6 } {\rm Re}[(G^{Z}_{ij})^2]
r_{j} \Bigr ){\phantom{xxxx}}
\eea
where $r_{Z} = m_{Z}/m_{\stilde N_i}$, and the neutralino-neutralino-$Z$
coupling is given by
\beq
G^{Z}_{ij} =
{g\over 2c_W} (-N_{i3} N^*_{j3} + N_{i4} N^*_{j4}) .
\label{GNSZ} 
\eeq

Finally we consider three-body decays $\stilde N_i \rightarrow f \overline
f \stilde N_j$, where $f$ is any Standard Model quark or lepton.
Results for these decays have appeared in refs. \cite{Franke,Bartl,BCDPT},
but here I
include the effects of Higgs boson exchanges and arbitrary phases of the
couplings .
The differential partial widths can be expressed in
terms of the dimensionless 
mass ratios $r_f = m_f/m_{\stilde N_i}$,
$r_{j} = m_{\stilde N_j}/m_{\stilde N_i}$,
$r_{Z} = m_Z/m_{\stilde N_i}$, 
$r_{\Gamma_Z} = \Gamma_Z/m_{\stilde N_i}$, and $r_\phi =
m_{\phi}/m_{\stilde N_i}$ for each of $\phi = h^0, A^0$, and $H^0$.
I use dimensionless kinematic variables
\beq
\hats &=& (p_{\stilde N_i} - p_{\stilde N_j})^2/p_{\stilde N_i}^2 
\\
\hatt &=& (p_{\stilde N_i} - p_{f})^2/p_{\stilde N_i}^2 
\\
\hatu &=& 1 + r_{j}^2 + 2 r_f^2 - \hats - \hatt
\eeq
with limits of integration
\beq
\phantom{x}& \phantom{x} &\hatt_{{\rm min},{\rm max}} = 
{1\over 2} [ 1 + r_{j}^2 - \hats + 2 r_f^2\mp \lbrace (1 - 4
r_f^2/\hats)\, \lambda(1,\hats ,r_{j}^2)\rbrace^{1/2}
] ;
\\
\phantom{x} & \phantom{x} & \hats_{\rm min} = 4 r_f^2;\qquad\qquad
\hats_{\rm max} \, =\,  (1 - r_{j})^2 .
\eeq
The results for widths can be expressed as\footnote{In computing these
results, I have
neglected fermion
masses arising from spinor algebra in
matrix elements, but not in the kinematic limits of integration or the
couplings.}
\beq
d\Gamma(\stilde N_i \rightarrow f \overline f \stilde N_j)
\,=\,
{n_c m_{\stilde N_i} \over 512 \pi^3} \, (\sum W) \, d\hat t d\hat s \, ,
\label{threebody}
\eeq
where $n_c = 1$ (3) for leptons (quarks). The individual contributions
to $\sum W$ are:
\beq
W_Z& \!=\! & {4 (a_f^2 + b_f^2)\over (\hats - r_Z^2)^2 +
r_Z^2
r_{\Gamma_Z}^2}
\Bigl \lbrace |G^Z_{ij}|^2 \left [ (1 - \hatu)(\hatu - r_{j}^2) +
(1 - \hatt)(\hatt -  r_{j}^2) \right ] 
\phantom{+ 2 {\rm Re}[(G^Z_{ij})^2]  r_{j} \hats \Bigr )xxxx}
\nonumber  \\   &&\qquad\qquad\qquad\qquad\qquad
+ 2 {\rm Re}[(G^Z_{ij})^2]  r_{j} \hats \Bigr \rbrace
\phantom{xxx}
\\
W_t &\!=\!& \sum_{n,n' = 1}^2
(a^{n}_{j} a^{n^\prime*}_{j} + b^n_{j} b_{j}^{n^\prime *})
(a^{n*}_{i} a^{n'}_{i} + b^{n*}_{i} b^{n^\prime}_{i} )
{(1 - \hatt )(\hatt - r_{j}^2) \over
(r_{\stilde f_n}^2 - \hatt)(r_{\stilde f_{n'}}^2 - \hatt)}
\\
W_u &\!=\!&  W_t(\hatt \rightarrow \hatu) \\
W_{tu} &\!=\!&  2 {\rm Re}\sum_{n,n'=1}^2 
{1\over
(r_{\stilde f_n}^2 - \hatt)(r_{\stilde f_{n'}}^2 - \hatu)}\Bigl [
(a^n_{j} b^{n^\prime}_{j} a^{n^\prime *}_{i} b^{n*}_{i} +
a^{n'}_{j} b^{n}_{j} a^{n *}_{i} b^{n'*}_{i}) 
(r_{j}^2 - \hatt \hatu ) \nonumber \\   
&& \>\>\qquad\qquad\qquad\qquad\qquad\qquad \>\> + 
(a^n_{j} a^{n^\prime}_{j} a_{i}^{n*} a_{i}^{n^\prime *} + b^n_{j}
b^{n^\prime}_{j} b_{i}^{n*} b_{i}^{n^\prime *})
\hats r_{j} \Bigr ]
\\
W_{Zt} &\!=\!&  {4 (\hats - r_Z^2 ) \over 
(\hats - r_Z^2)^2 + r_Z^2 r^2_{\Gamma_Z}}
{\rm Re}\sum_{n=1}^2 
\Bigl [
(a_f a^n_i a^{n*}_j + b_f b^{n*}_i b^{n}_j ) 
\lbrace G_{ij}^Z (1 - \hatt)(\hatt - r_{j}^2) \nonumber \\
&& \qquad\qquad\qquad\qquad\qquad\qquad\qquad
\qquad\>\>\>\>\>
+ G^{Z*}_{ij} \hats r_{j}\rbrace 
\Bigr ]/(r_{\stilde f_n}^2 - \hatt) 
\\
W_{Zu} &=& W_{Zt} (\hatt \rightarrow \hatu)
\\
W_{h^0,H^0} &=& \sum_{\phi,\phi' = h^0, H^0}
{4 \hats \,{\rm Re}[G_f^\phi G_f^{\phi' *}] \over
(r_\phi^2 -\hats)
(r_{\phi'}^2 -\hats)} \left \lbrace
(1 + r_j^2 - \hats) {\rm Re}[G_{ij}^\phi G_{ij}^{\phi' *}] + 2 \hats r_j
{\rm Re}[G_{ij}^\phi G_{ij}^{\phi'}]
\right \rbrace
\\
W_{A^0} &=& {4 \hats
|G_f^{A^0}|^2 \over (r_{A^0}^2 -\hats )^2} \left \lbrace
(1 + r_j^2 - \hats) |G^{A_0}_{ij}|^2 + 2 \hats r_j
{\rm Re}[(G^{A_0}_{ij})^2 ]
\right \rbrace
\\
W_{\phi t} &=& - \sum_{\phi = h^0,H^0,A^0}\sum_{n=1,2}
{2\over (r_\phi^2 - \hats)(r_{\stilde f_n}^2 - \hatt)}
{\rm Re}[\hats G^\phi_f 
(\hatt G^\phi_{ij} + r_j G^{\phi *}_{ij}) 
(a_i^{n*} b_j^n + a_j^{n*}b^n_i) ] 
\\
W_{\phi u} &=& W_{\phi t}(\hatt \rightarrow \hatu) .
\eeq
Quantities appearing in the above results are as follows. First,
\beq 
a_f = -{g\over c_W}( T_{3f} - q_f s_W^2);\qquad\qquad
b_f = - q_f g s_W^2/c_W
\eeq
are the $Z$ boson couplings to quarks and leptons with $(T_{3f},q_f) =
(1/2,2/3)$ for up-type quarks, $(-1/2,-1/3)$ for down-type quarks,
and $(-1/2,-1)$ for charged leptons.
Left-right mixing and CP violation in
the
sfermion sector are parameterized by a unitary matrix, which I
choose\footnote{This parameterization has the feature that
in the typical unmixed, CP-conserving case $c_{\stilde f}= 1$, $s_{\stilde
f} = 0$,
$\stilde f_1 = \stilde f_R$ and $\stilde f_2 = \stilde f_L$, with no minus
signs.}
to write as
\beq
\pmatrix{\stilde f_R \cr \stilde f_L} = \pmatrix{
c_{\stilde f} & s_{\stilde f} \cr
-s^*_{\stilde f} & c^*_{\stilde f}} \pmatrix{\stilde f_1 \cr \stilde f_2}
, 
\label{smixing}
\eeq
where $|c_{\stilde f}|^2 + |s_{\stilde f}|^2 = 1$,
and
$m_{\stilde f_1} < m_{\stilde f_2}$,
The resulting couplings for down-type sfermions $(\stilde
b,\stilde \tau)$ are: 
\beq
a^{\stilde f_1}_{i} &=& \sqrt{2} s_{\stilde f} 
\left [ g T_{3f} N_{i2}^* + g' (q_f - T_{3f}) N_{i1}^* \right ]
- {c^*_{\stilde f} g N_{i3}^* m_f/ \sqrt{2} c_\beta m_W} ,
\label{defaco}
\\
b^{\stilde f_1}_{i} &=& \sqrt{2} c^*_{\stilde f} g' q_f N_{i1}
+ {s_{\stilde f} g N_{i3} m_f/ \sqrt{2} c_{\beta} m_W}
\label{defbco} \\
a^{\stilde f_2}_{i} &=& - \sqrt{2} c_{\stilde f} 
\left [ g T_{3f} N_{i2}^* + g' (q_f - T_{3f}) N_{i1}^* \right ]
- {s^*_{\stilde f} g N_{i3}^* m_f/ \sqrt{2} c_\beta m_W} ,
\label{defacot}
\\
b^{\stilde f_2}_{i} &=& \sqrt{2} s^*_{\stilde f} g' q_f N_{i1}
- {c_{\stilde f} g N_{i3} m_f/ \sqrt{2} c_{\beta} m_W}
\label{defbcot}
\eeq
for $i=0,1,\ldots 4$. 
For up-type fermions one must replace $N^{(*)}_{i3}/c_\beta$ by
$N^{(*)}_{i4} /s_\beta$ in eqs.~(\ref{defaco})-(\ref{defbcot}).
(However, for the cases of interest in this paper,
$\stilde N_1 \rightarrow t \overline t \stilde S$ is
surely not
kinematically allowed, and decays $\stilde N_1 \rightarrow 
\nu\overline \nu S$ are not interesting.)
In the expressions for the contributions to the widths, I have used the
abbreviations
$a_{i}^{n} = a_i^{\stilde f_n}$, etc. 
The various Higgs boson couplings to Standard Model fermions
are given by, e.g.:
\beq
&&G_b^{h^0} = {g m_b s_\alpha \over 2 m_W
c_\beta};\qquad\qquad\>\>\>\>\>\>\>\>\>\>
G_t^{h^0} = -{g m_t c_\alpha \over 2 m_W s_\beta};\\
&&G_b^{H^0} = -{g m_b c_\alpha \over 2 m_W c_\beta};\qquad\qquad\>\>\>\>\>
G_t^{H^0} = -{g m_t s_\alpha \over 2 m_W s_\beta};\\
&&G_b^{A^0} = -i{g m_b \tan\beta \over 2 m_W };\qquad\qquad\>\>
G_t^{A^0} = -i{g m_t \cot\beta \over 2 m_W},
\eeq
for bottom and top quarks.
The Higgs couplings for taus are obtained by $m_b
\rightarrow m_\tau$, and for other quarks and leptons by the obvious
substitutions.

The two-body decay width for a stau to a neutralino or singlino is
given by:
\bea
\Gamma(\stilde \tau_1 \rightarrow \tau \stilde N_i) &\! = \! &
{m_{\stilde \tau_1} \over 16 \pi} \sqrt{\lambda ( 1, {
r_{i}^2}, {r_{\tau }^2} )}
\Bigl \lbrace
(|a^{\stilde \tau_1}_i|^2 + |b^{\stilde \tau_1}_i|^2) (1 - 
{r_{i}^2}- 
{r_{\tau}^2}) 
- {4 r_\tau r_{i}} {\rm Re}
[a^{\stilde \tau_1}_i b^{\stilde \tau_1 *}_i ]
\Bigr \rbrace {\phantom{xxxx}}
\eea
where now $r_{i} = m_{\stilde N_i}/m_{\stilde \tau_1}$ and 
$r_{\tau} = m_\tau/m_{\stilde \tau_1}$, and the couplings $a^{\stilde
\tau_1}_i$, $b^{\stilde \tau_1}_i$ are given already by
eqs.~(\ref{defaco})-(\ref{defbcot}). 
In section 4, this formula is used with $i=0$, corresponding to $\stilde
\tau_1 \rightarrow \tau \stilde S $.
The results for $\stilde \mu_R
\rightarrow \mu \stilde S$ and
$\stilde e_R \rightarrow e \stilde S$ are obtained by taking $c_{\stilde
\tau} \rightarrow 1$ and $s_{\stilde \tau} \rightarrow 0$.


\end{document}